\documentclass[12pt,a4paper]{article}
\usepackage{graphicx}
\usepackage{amsbsy} 
\newcommand{\vs}{\vspace{-0.25cm}}

\begin{document}
\begin{center}
{\Large
\textbf{Isovector part of nuclear energy density functional \\ 
from chiral two- and three-nucleon forces}\footnote{Work supported in part by 
BMBF, GSI and the DFG
cluster of excellence: Origin and Structure of the Universe.}} 

\bigskip
N. Kaiser\\

\bigskip

{\small Physik Department T39, Technische Universit\"{a}t M\"{u}nchen, 
D-85747 Garching, Germany\\

\smallskip

{\it email: nkaiser@ph.tum.de}}

\end{center}

\begin{abstract}
A recent calculation of the nuclear energy density functional from chiral two- and 
three-nucleon forces is extended to the isovector terms pertaining to different 
proton and neutron densities. An improved density-matrix expansion is 
adapted to the situation of small isospin-asymmetries and used to calculate 
in the Hartree-Fock approximation the density-dependent strength functions associated 
with the isovector terms. The two-body interaction comprises of long-range 
multi-pion exchange contributions and a set of contact terms contributing 
up to fourth power in momenta. In addition, the leading order chiral three-nucleon 
interaction is employed with its parameters fixed in computations of nuclear 
few-body systems. With this input one finds for the asymmetry energy of nuclear 
matter the value  $A(\rho_0) \simeq 26.5\,$MeV, compatible with existing 
semi-empirical determinations. The strength functions of the isovector surface and 
spin-orbit coupling terms come out much smaller than those of the analogous 
isoscalar coupling terms and in the relevant density range one finds agreement with 
phenomenological Skyrme forces. The specific isospin- and density-dependences arising 
from the chiral two- and three-nucleon interactions can be explored and tested 
in neutron-rich systems. 
\end{abstract}

\smallskip


\section{Introduction}
The nuclear energy density functional approach is the many-body method of choice in 
order to calculate the properties of medium-mass and heavy nuclei in a systematic 
manner \cite{reinhard,stone}. Parameterized non-relativistic Skyrme functionals 
\cite{sly,pearson} as well as relativistic mean-field models \cite{serot,ring} have 
been widely and successfully used for such nuclear structure calculations. In a  
complementary approach one attempts to constrain the analytical form of the 
functional and the values of its couplings from many-body perturbation theory and 
the underlying two- and three-nucleon interaction. Switching from conventional 
hard-core NN-potentials to low-momentum interactions \cite{vlowkreview,vlowk} is 
essential in this respect, because the nuclear many-body problem formulated in terms 
of the latter becomes significantly more perturbative. 

In many-body perturbation theory the contributions to the energy are written in 
terms of density-matrices convoluted with the finite-range interaction kernels, and 
are therefore highly non-local in both space and time. In order to make such 
functionals numerically tractable in heavy open-shell nuclei it is necessary to 
develop simplified approximations for these functionals in terms of local densities 
and currents. In such a construction the density-matrix expansion comes 
prominently into play as it removes the non-local character of the exchange (Fock) 
contribution to the energy by mapping it onto a generalized Skyrme functional with 
density-dependent couplings. For some time the prototype for that has been the 
density-matrix expansion of Negele and Vautherin \cite{negele}, but recently 
Gebremariam, Duguet and Bogner \cite{dmeimprov} have developed an improved version 
for spin-unsaturated nuclei. They have demonstrated that phase-space averaging 
techniques allow for a consistent expansion of both the spin-independent (scalar) 
part as well as the spin-dependent (vector) part of the density-matrix. 

By applying these new techniques a microscopically constrained nuclear 
energy density functional has been derived from the chiral NN-potential at 
next-to-next-to-leading order (N$^2$LO) in ref.\cite{microefun} by Gebremariam, 
Bogner and Duguet. These authors have proposed that the density-dependent couplings 
associated with the pion-exchange interactions should be added to a standard Skyrme 
functional (with several adjustable parameters). In the sequel it has been 
demonstrated in ref.\cite{stoitsov} that this new energy density functional gives 
numerically stable results and that it exhibits a small but systematic reduction of 
the $\chi^2$-measure compared to standard Skyrme functionals (without any 
pion-exchange terms).

In the recent work \cite{chiral23fun} the calculation of the nuclear energy 
density functional has been continued and extended with improved (chiral) two- and 
three-nucleon interactions as input. For the two-body interaction the N$^3$LO chiral 
NN-potential has been used in ref.\cite{chiral23fun}. It consists  
of long-range multi-pion exchange terms and two dozen low-energy constants which 
parameterize the short-distance part of the NN-interaction. The actual 
calculation in ref.\cite{chiral23fun} has been performed with the version 
N$^3$LOW developed in ref.\cite{n3low} by lowering the cut-off scale to 
$\Lambda= 414\,$MeV. This value coincides with the resolution scale below which 
evolved low-momentum NN-potentials become nearly model-independent and exhibit 
desirable convergence properties in perturbative many-body  calculations 
\cite{vlowkreview,vlowk,achim}. The (low-momentum) two-body interaction N$^3$LOW has 
been supplemented in ref.\cite{chiral23fun} by the leading order (N$^2$LO) chiral 
three-nucleon interaction with its parameters $c_E$, $c_D$ and $c_{1,3,4}$ determined 
in computations of nuclear few-body systems \cite{achim,3bodycalc}. With this input 
the nuclear energy density functional has been derived to first order in many-body 
perturbation theory, i.e. in the Hartree-Fock approximation. For the effective 
nucleon mass $M^*(\rho)$ and the strength functions $F_\nabla(\rho)$ and $F_{so}(\rho)$
 of the (isoscalar) surface and 
spin-orbit coupling terms reasonable agreement with results of phenomenological 
Skyrme forces has been found (in the relevant density range).  However, as indicated 
in particular by the nuclear matter equation of state $\bar E(\rho)$, an improved 
description of the energy density functional requires at least the treatment of the 
two-nucleon interaction to second order in many-body perturbation theory. 

The purpose of the present paper is to extend the calculation of the nuclear energy 
density functional in ref.\cite{chiral23fun} to isospin-asymmetric many-nucleon 
systems with different proton and neutron densities. The additional isovector terms 
play an important role in the description of long chains of stable isotopes and for 
nuclei far from stability. Our paper is organized as follows. In section 2 we recall 
the improved density-matrix expansion  of Gebremariam, Duguet and Bogner 
\cite{dmeimprov} whose Fourier transform to momentum space provides the adequate 
technical tool to calculate the nuclear energy density functional in a diagrammatic 
framework. In section 3 we present the two-body contributions to the various
density-dependent strength functions $\tilde A(\rho)$, $G_\tau(\rho)$,  $G_d(\rho)$, 
$G_{so}(\rho)$ and $G_J(\rho)$, separately for the finite-range pion-exchange and the 
zero-range contact interactions. Section 4 comprises the corresponding 
analytical expressions for the three-body contributions grouped into contact 
$(c_E)$, $1\pi$-exchange ($c_D$) and  $2\pi$-exchange ($c_{1,3,4}$) terms. Finally,  we 
discuss in section 5 our numerical results and add some concluding remarks.
\section{Density-matrix expansion and isovector part of energy 
density functional}
The starting point for the construction of an explicit nuclear energy density 
functional is the bilocal density-matrix as given by a sum over the orbitals  
occupied by protons and neutrons: $\sum_\alpha\!\Psi^{(\alpha)}_{p,n}(\vec r-\vec a/2)
\Psi^{(\alpha)\dagger}_{p,n}(\vec r+\vec a/2)$. According to  Gebremariam, Duguet and 
Bogner \cite{dmeimprov} it can be expanded in relative and center-of-mass 
coordinates, $\vec a$ and $\vec r$, with expansion coefficients determined by local 
proton and neutron densities. These are the particle densities $\rho_{p,n}(\vec r\,)$, 
the kinetic energy densities $\tau_{p,n}(\vec r\,)$ and the spin-orbit densities  
$\vec J_{p,n}(\vec r\,)$ (for definitions in terms of the orbitals $\Psi^{(\alpha)}_{p,n}
(\vec r\,)$, see section 2 in ref.\cite{chiral23fun}). The Fourier transform of the 
expanded density-matrix with respect to both coordinates defines in momentum space a 
medium insertion: 
\begin{eqnarray} \Gamma(\vec p,\vec q\,)& =& \int\!d^3 r \, e^{-i \vec q \cdot
\vec r}\,\bigg\{ {{\boldsymbol 1}+{\boldsymbol\tau}_3\over 2}\,\theta(k_p-
|\vec p\,|) + {{\boldsymbol 1}-{\boldsymbol\tau}_3\over 2}\,
\theta(k_n-|\vec p\,|)  \nonumber \\ && +{\pi^2 \over 4k_f^4}\Big[k_f\,\delta'
(k_f-|\vec p\,|)-2 \delta(k_f-|\vec p\,|)\Big] \bigg[ \tau_p - \tau_n-\bigg(
k_f^2+{\vec \nabla^2 \over 4}\bigg)\nonumber \\ && \times (\rho_p-\rho_n)\bigg] 
{\boldsymbol\tau}_3-{3\pi^2 \over 4k_f^4}\,\delta(k_f-|\vec p\,|) \, (\vec 
\sigma \times\vec p\,) \cdot( \vec J_p- \vec J_n) {\boldsymbol \tau}_3+\dots 
\bigg\}\,,  \end{eqnarray}
for the inhomogeneous isospin-asymmetric many-nucleon system. Here,  
${\boldsymbol\tau}_3$ denotes the third Pauli isospin-matrix and we have displayed 
only the (relevant) terms proportional to differences of proton and neutron 
densities: $\rho_p - \rho_n$, $\tau_p - \tau_n$, $\vec J_p- \vec J_n$. The local 
Fermi momenta $k_{p,n,f}(\vec r\,)$ are related to the (particle) densities in the 
usual way: $\rho_p=k_p^3/3\pi^2$, $\rho_n=k_n^3/3\pi^2$, $\rho=\rho_p+\rho_n=2k_f^3 
/3\pi^2$. When working to quadratic order in deviations from isospin symmetry (i.e. 
proton-neutron differences) it is sufficient to use an average Fermi momentum $k_f$ 
in the prefactors of  $\tau_p - \tau_n$ and $\vec J_p- \vec J_n$. 

Up to second order in proton-neutron differences and spatial gradients the isovector 
part of the nuclear energy density functional takes the form:
\begin{eqnarray} &&{\cal E}_{\rm iv}[\rho_p,\rho_n,\tau_p,\tau_n,\vec J_p,\vec J_n] 
= {1\over \rho}(\rho_p-\rho_n)^2\,\tilde A(\rho)+ {1\over \rho}(\tau_p-\tau_n)
(\rho_p-\rho_n)\, G_\tau(\rho) \nonumber \\ && + (\vec \nabla \rho_p-\vec \nabla 
\rho_n)^2\, G_\nabla(\rho)+  (\vec \nabla \rho_p- \vec \nabla \rho_n)\cdot(\vec J_p-
\vec J_n)\, G_{so}(\rho)+ (\vec J_p-\vec J_n)^2 \, G_J(\rho)\,.\end{eqnarray}
Here, $\tilde A(\rho)$ is the interacting part of the asymmetry energy of 
(homogeneous) nuclear matter. The non-interacting (kinetic energy) contribution 
$A_{\rm kin}(\rho) = k_f^2/6M$ to the asymmetry energy is included in the nuclear 
energy density functional through the kinetic energy density term, $ {\cal E}_{\rm kin}
=(\tau_p+\tau_n)/2M$, with $M=939\,$MeV the (free) nucleon mass.  The strength 
function $G_\nabla(\rho)$ of the isovector surface term $(\vec \nabla \rho_p-\vec 
\nabla \rho_n)^2$ has the decomposition:
\begin{equation} G_\nabla(\rho) = {1\over 4\rho}\, G_\tau(\rho)+G_d(\rho) \,,
\end{equation}
where $G_d(\rho)$ comprises all those contributions for which the $(\vec \nabla 
\rho_p-\vec \nabla \rho_n)^2$  factor originates directly from the momentum 
dependence of the interactions in an expansion up to order $\vec q^{\,2}$. The 
Fourier transformation in eq.(1) converts this factor  $\vec q^{\,2}$ into 
$(\vec \nabla k_p-\vec \nabla k_n)^2 \simeq (\vec \nabla \rho_p-\vec \nabla \rho_n)^2 
(\pi/k_f)^4$. The second last term $(\vec \nabla \rho_p- \vec \nabla \rho_n)\cdot
(\vec J_p-\vec J_n)\, G_{so}(\rho)$ in eq.(2) describes the isovector spin-orbit 
interaction in nuclei. Depending on the sign and size of its strength function 
$G_{so}(\rho)$ the spin-orbit potentials for protons and neutrons are differently
composed from the gradients of the local proton and neutron densities.   

\section{Two-body contributions}
In this section the two-body contributions to the various strength functions 
$\tilde A(\rho)$, $G_\tau(\rho)$,  $G_d(\rho)$, $G_{so}(\rho)$ and $G_J(\rho)$ are 
worked out. We follow closely section 3 in ref.\cite{chiral23fun} where the input 
two-body interaction, the chiral nucleon-nucleon potential N$^3$LOW \cite{n3low}, 
has been described in sufficient detail. In the (first-order) Hartree-Fock 
approximation the finite-range multi-pion exchange interactions lead in combination 
with the density-matrix expansion (i.e. by employing the product of two medium 
insertions $\Gamma(\vec p_1,\vec q\,)\,\Gamma(\vec p_2,-\vec q\,)$) to the 
following two-body contributions to the strength functions:
\begin{eqnarray} \tilde A(\rho) &=& {\rho\over 2} W_C(0) - {\rho\over 2} \int_0^1
\!\!dx\,\bigg\{ x^3\Big[V_C(q)+3V_S(q)+q^2 V_T(q)\Big] \nonumber \\ && \qquad
\qquad\quad +(3x^3-2x)\Big[W_C(q)+3W_S(q)+q^2 W_T(q)\Big] \bigg\}\,, \end{eqnarray}
\begin{equation} G_\tau(\rho) = {k_f\over 6\pi^2}\bigg\{-{1\over 2}U(2k_f)+
\int_0^1\!\!dx\,x\,U(2xk_f)\bigg\}\,, \end{equation}
with the (isoscalar minus isovector) combination of the central, spin-spin and 
tensor NN-potentials in momentum space: 
\begin{equation}U(q)=V_C(q)-W_C(q) +3V_S(q)-3W_S(q) +q^2V_T(q) -q^2W_T(q) \,.
\end{equation}
\begin{equation} G_d(\rho) = {1\over 4} W_C''(0)\simeq  -9.9\,{\rm MeV fm}^5 \,,
\end{equation} 
\begin{equation} G_{so}(\rho) = {1\over 2}W_{SO}(0)+ \int_0^1\!\!dx\,x^3\Big[
V_{SO}(2x k_f)-W_{SO}(2x k_f)\Big] \,, \end{equation}
\begin{equation} G_J(\rho) = {3\over 8k_f^2} \int_0^1\!\!dx\bigg\{(2x^3-x)\Big[
V_C(q)-W_C(q) -V_S(q)+W_S(q)\Big] +x^3 q^2\Big[W_T(q)-V_T(q)\Big]\bigg\}
\,. \end{equation} 
In the integrands of eqs.(4,9) the  momentum transfer variable $q$ is to be set to 
$q = 2x k_f$. The double-prime in eq.(7) denotes a second derivative and we have 
given the numerical value for $G_d(\rho)$ resulting from the (negative) curvature of 
the isovector central potential $W_C(q)$ shown in Fig.\,1 of ref.\cite{chiral23fun}.

In addition there are the two-body contributions from the zero-range contact 
potential of the chiral NN-interaction N$^3$LOW. The  corresponding expression in 
momentum space includes constant, quadratic, and quartic terms in momenta and it can 
be found in section 2.2 of ref.\cite{evgeni}. The Hartree-Fock contributions from 
the NN-contact potential to the strength functions read:
\begin{equation} \tilde A(\rho) = -{\rho\over 8}(C_S+3C_T)+{\rho k_f^2 \over 12} 
(C_2-4C_1-12C_3-4C_6)+\rho k_f^4\bigg({D_2\over 12}-D_1-3D_5-D_{11}\bigg)\,,
\end{equation} 
\begin{equation} G_\tau(\rho) = -{\rho \over 4} (C_1+3C_3+C_6)-{4\rho k_f^2\over
 3}(D_1+3D_5+D_{11})\,,\end{equation} 
\begin{equation} G_d(\rho) = -{1\over 32} (C_2+3C_4+C_7)-{k_f^2\over 48}
(3D_3+2D_4+9D_7+6D_8+3D_{12}+3D_{13}+2D_{15})\,,\end{equation} 
\begin{equation} G_{so}(\rho) = {C_5\over 8} +{k_f^2\over 3} D_9\,,
\end{equation}
\begin{equation} G_J(\rho) = {1\over 8} (C_1-C_3-2C_6)+{k_f^2\over 4}
(2D_1-2D_5-3D_{11})\,.\end{equation} 
The 24 low-energy constants $C_{S,T}$, $C_j$ and $D_j$ are determined (at the cut-off 
scale of $\Lambda = 414\,$MeV) in fits to empirical NN-phase shifts and deuteron 
properties \cite{n3low}. Their numerical values have been extracted from the 
pertinent NN-scattering code and are listed in section 3 of ref.\cite{chiral23fun}.
Let us mention that the contributions proportional to $C_{S,T}$ and $C_j$ in 
eqs.(10-14) have also been worked out in appendix B of ref.\cite{microefun} and we 
find agreement with their results. The terms proportional to $D_j$ as well as the 
master formulas eqs.(4-9) for the finite-range contributions are new.
\section{Three-body contributions}
In this section the three-body contributions to the  strength functions $\tilde 
A(\rho)$, $G_\tau(\rho)$, $G_d(\rho)$, $G_{so}(\rho)$ and  $G_J(\rho)$ are worked out. 
We employ the leading order chiral three-nucleon interaction \cite{3bodycalc} which 
consists of a contact piece (with parameter $c_E$), a $1\pi$-exchange component (with 
parameter $c_D$) and a $2\pi$-exchange component (with parameters $c_1$, $c_3$ and 
$c_4$). In order to treat the three-body correlations in isospin-asymmetric 
inhomogeneous nuclear many-body systems we assume (as done in ref.\cite{chiral23fun}) 
that the relevant product of density-matrices can be represented in momentum space in 
a factorized form by $\Gamma(\vec p_1,\vec q_1)\,\Gamma(\vec p_2,\vec q_2)\,
\Gamma(\vec p_3,-\vec q_1-\vec q_2)$. Such a factorization ansatz respects by 
construction the correct nuclear matter limit, but it involves approximations in 
comparison to more sophisticated treatments outlined in section 4 of 
ref.\cite{platter}. Actually, the present approach is similar to the method DME-I 
introduced in ref.\cite{platter}. In comparison to ref.\cite{chiral23fun} the 
diagrammatic calculation of the isovector terms gets essentially modified only by 
relative isospin factors occurring at various places. However, their pattern is rather 
complex and therefore it is preferable to write out each (non-vanishing) contribution 
individually. We give for each diagram only the final result omitting all technical 
details related to extensive algebraic manipulations, expansions, and solving 
elementary integrals.      
\subsection{$c_E$-term}
The three-body contribution from the contact interaction is represented by the  
left diagram in Fig.\,1. One finds a contribution to the asymmetry energy:  
\begin{equation} \tilde A(\rho) = {3c_E \rho^2 \over 16f_\pi^4\Lambda_\chi}\,,
\end{equation}
which depends quadratically on the density $\rho=2k_f^3/3\pi^2$ and is equal with 
opposite sign to the contribution to the energy per particle $\bar E(\rho)$. This 
property follows from the form $\rho_p \rho_n(\rho_p+\rho_n)$ of the underlying 
energy density as it is determined by the Pauli exclusion principle and the symmetry 
under $p\leftrightarrow n$ exchange.  Due to the momentum-independence of the 
three-body contact interaction the contributions to the other strength functions 
$G_{\tau,d,so,J}(\rho)$ vanish.

\begin{figure}
\begin{center}
\includegraphics[scale=1.2,clip]{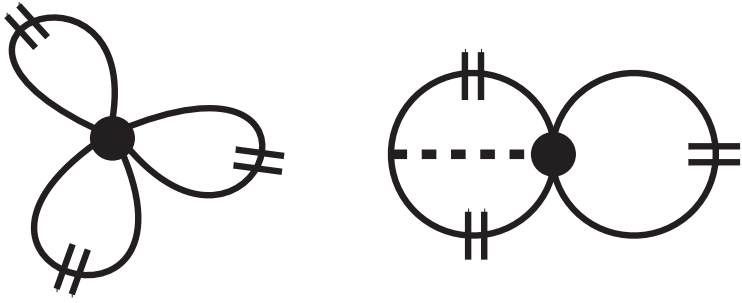}
\end{center}
\vspace{-.6cm}
\caption{Three-body diagrams related to the contact ($c_E$) and $1\pi$-exchange 
($c_D$) component of the chiral three-nucleon interaction. The short double-line 
symbolizes the medium insertion $\Gamma(\vec p, \vec q\,)$ for inhomogeneous 
isospin-asymmetric nuclear matter.}
\end{figure}

\subsection{$c_D$-term}
Next, we consider the three-body contributions from the $1\pi$-exchange component of 
the chiral 3N-interaction as represented by the right diagram in Fig.\,1. Putting in 
three medium insertions one finds the following analytical expressions:  
\begin{equation} \tilde A(\rho)={g_A c_D m_\pi^6 u^2\over(2\pi f_\pi)^4\Lambda_\chi}
\bigg\{{u^2 \over 6}-{u^4\over 3}-{u\over 3}\arctan 2u +\bigg({1\over 8}+
{u^2 \over 9}\bigg)\ln(1+4u^2) \bigg\}\,,\end{equation}  
\begin{equation} G_\tau(\rho)={g_A c_D m_\pi^4\over 18(2\pi f_\pi)^4\Lambda_\chi}
\bigg\{ {3u^2+14u^4 \over 1+4u^2}-\bigg({3\over 4}+2u^2\bigg) \ln(1+4u^2)
\bigg\}\,,\end{equation}
\begin{equation} G_d(\rho)={g_A c_D m_\pi\over(4f_\pi)^4 \pi^2\Lambda_\chi}
\bigg\{{2u\over 3(1+4u^2)}- {1\over 6u}\ln(1+4u^2) \bigg\}\,,\end{equation}
\begin{equation} G_J(\rho)={g_A c_D m_\pi\over(4f_\pi)^4 \pi^2\Lambda_\chi}
\bigg\{ {1\over u}-2u -{1\over 4u^3}\ln(1+4u^2)\bigg\}\,, \end{equation}
with the abbreviation $u=k_f/m_\pi$. Note that there is no contribution to 
the isovector spin-orbit coupling strength $G_{so}(\rho)$, essentially because the 
$1\pi$-exchange does not generate any.   

\subsection{Hartree diagram proportional to $c_{1,3}$}
\begin{figure}
\begin{center}
\includegraphics[scale=1.2,clip]{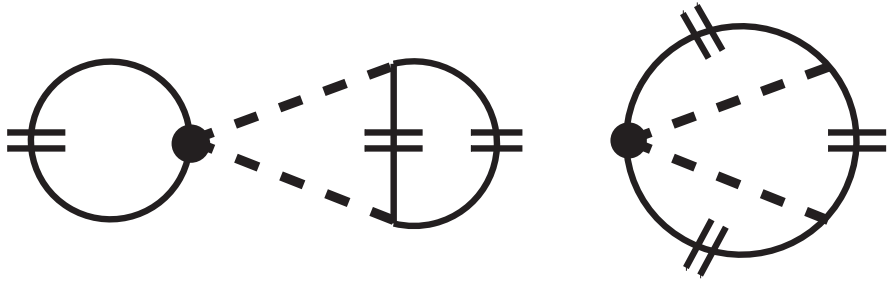}
\end{center}
\vspace{-.6cm}
\caption{Three-body Hartree and Fock diagrams related to the chiral $2\pi$-exchange 
three-nucleon interaction.}
\end{figure}

We continue with the three-body contributions from the $2\pi$-exchange Hartree diagram 
shown in the left part of Fig.\,2. Again putting in three medium insertions one 
derives the following analytical results:
\begin{eqnarray}  \tilde A(\rho)&=&{2g_A^2 m_\pi^6 u^2\over 9(2\pi f_\pi)^4}
\bigg\{{(c_3-2c_1)u^2 \over 1+4u^2} +(8c_3-10c_1)u^2 +2c_3 u^4\nonumber \\ && 
+ \bigg[3c_1- {9c_3 \over 4} +4(c_1-c_3)u^2\bigg]\ln(1+4u^2)\bigg\}\,,  
\end{eqnarray}
\begin{equation}  G_\tau(\rho) = {2g_A^2 m_\pi^4 u^2\over 9(2\pi f_\pi)^4}\bigg\{
(c_3-c_1)\ln(1+4u^2) +{4u^2 \over (1+4u^2)^2}\Big[c_1-c_3+(8c_1-6c_3)u^2 \Big]
\bigg\}\,,  \end{equation}
\begin{equation} G_J(\rho)  = {g_A^2 m_\pi \over (8\pi)^2 f_\pi^4}\bigg\{
{4c_1-3c_3 \over u} +2c_3 u+{4u(c_3-2c_1)\over 1+4u^2}+{3c_3-4c_1 \over 4u^3}
\ln(1+4u^2) \bigg\}\,, \end{equation}
which depend only on the two isoscalar coupling constants $c_1$ and $c_3$. The 
isovectorial (spin-dependent) $c_4$-vertex gets eliminated by a vanishing spin-trace 
(over the left nucleon ring). The vanishing contributions to $G_d(\rho)$ and  
$G_{so}(\rho)$ from the $2\pi$-exchange three-body Hartree diagram are particularly 
remarkable, in view of the fact that their isoscalar counterparts ($F_d(\rho)$ and 
$F_{so}(\rho)$ in eqs.(24,25) of ref.\cite{chiral23fun}) are quite sizeable. The 
actual calculation shows that the isospin-structure of the $c_{1,3}$-vertex excludes 
the desired coupling of the gradient $\vec\nabla k_p- \vec\nabla k_n$ to the vectors 
$\vec J_p- \vec J_n$ and $\vec\nabla k_p- \vec\nabla k_n$. 

\subsection{Fock diagram proportional to $c_{1,3,4}$}
Finally, there are the three-body contributions from the $2\pi$-exchange Fock diagram 
shown in the right part of Fig.\,2. For this diagram the occurring integrals over 
three Fermi spheres cannot be solved analytically in all cases. After a somewhat tedious
calculation of the separate pieces proportional to $c_1$, $c_3$ and $c_4$ one finds 
the following results for the Fock contributions to the strength functions: 
\begin{eqnarray}  \tilde A(\rho)&=&{g_A^2 m_\pi^6 \over 9(4\pi f_\pi)^4 u^3}
\int_0^u\!\!dx \bigg\{ 3c_1\Big[3H_{10}^2+3H_{01}^2-2H_{10}H_{01} +H\Big(3H_{20}+3H_{02} 
\nonumber \\ && -2H_{11}-8H_{01}-3H\Big)\Big]+\Big(c_4+{3c_3\over2}\Big) G_{S01}^2+
(2c_4-c_3)G_{S01}G_{S10}\nonumber \\ && +3\Big({c_3\over2}-c_4\Big) G_{S10}^2
+\Big(c_4-{c_3\over2}\Big)G_S\Big(3G_S+8 G_{S01}-3G_{S02}+2G_{S11}\nonumber \\ &&
-3G_{S20}\Big) +(3c_3-c_4)G_{T01}^2 -2(c_3+c_4)G_{T01}G_{T10}+3(c_3+c_4)G_{T10}^2 
\nonumber \\ && +(c_3+c_4)G_T\Big(3G_{T02}+3G_{T20}-2G_{T11}-8G_{T01}-3G_T\Big) 
\bigg\}\,, \end{eqnarray}
with the auxiliary functions:
\begin{equation} H(x,u) = u(1+x^2+u^2)-{1\over 4x}\Big[1+(u+x)^2\Big]\Big[1+(u-x)^2
\Big] \ln{1+(u+x)^2\over  1+(u-x)^2} \,,\end{equation}
\begin{eqnarray} G_S(x,u) &=& {4ux \over 3}( 2u^2-3) +4x\Big[
\arctan(u+x)+\arctan(u-x)\Big] \nonumber \\ && + (x^2-u^2-1) \ln{1+(u+x)^2
\over  1+(u-x)^2} \,,\\ G_T(x,u) &=& {ux\over 6}(8u^2+3x^2)-{u\over
2x} (1+u^2)^2  \nonumber \\ && + {1\over 8} \bigg[ {(1+u^2)^3 \over x^2} -x^4 
+(1-3u^2)(1+u^2-x^2)\bigg] \ln{1+(u+x)^2\over  1+(u-x)^2} \,.\end{eqnarray}
A double-index notation has been introduced for partial derivatives multiplied by 
powers of the variables $x$ and $u$: 
\begin{equation} H_{ij}(x,u) = x^i u^j \,{\partial^{i+j} H(x,u) \over \partial x^i 
\partial u^j} \,,\end{equation}
which applies in the same way to the functions $G_{Sij}(x,u)$ and $G_{Tij}(x,u)$.
\begin{eqnarray}  G_\tau(\rho) &=& {g_A^2 c_1 m_\pi^4 \over (2\pi f_\pi)^4} \Bigg\{
{7u^2 \over 6}+{5+16u^2 \over 12(1+4u^2)}-u \arctan 2u -{5+7u^2 \over 24u^2}
\ln(1+4u^2) \nonumber \\ && +{5+16u^2 \over 192u^4}\ln^2(1+4u^2) +\int_0^u\!\!dx 
\bigg\{{L^2 \over u}\Big[u^4-(1-x^2)^2\Big] \nonumber \\ && +2L\bigg[1-u^2 -{x(u+x) 
\over 1+(u+x)^2}+{x(u-x) \over 1+(u-x)^2} \bigg] \bigg\} \Bigg\}  \nonumber \\ &+& 
{g_A^2 c_3 m_\pi^4 \over (4\pi f_\pi)^4} \Bigg\{{7\over 6u^2}-{761 u^2 \over 54}
-{256u^4 \over 9}+{21+4u^2 \over 27(1+4u^2)} \nonumber \\ && +\bigg[ 10u +12u^3 
+{32u \over 9(1+4u^2)} - {8 \over 9u} \ln(1+4u^2)\bigg] \arctan 2u   \nonumber \\ && 
+\bigg( {83 \over 72}-{7\over 12u^4}-{14 \over 9u^2}-{37u^2 \over 54} -{8 \over 
9(1+4u^2)}\bigg)\ln(1+4u^2)\nonumber \\ && +\bigg({1\over 3}+{2\over 3u^2}+{49 \over 
144u^4}+{7 \over 96u^6} \bigg)\ln^2(1+4u^2) +\int_0^u\!\!dx \bigg\{{L^2 \over u}
\bigg[{3\over x^2}\nonumber \\ && \times(1+u^2)^3(3u^2-1)+4(1-7u^4-6u^6)+18x^2
(u^4-1)+4x^4-3x^6 \bigg]\nonumber \\ && +2L \bigg[7(2u^2-1+3u^4)+{3\over x^2}
(1+u^2)^2(1-3u^2) \nonumber \\ && +{8x(u+x) \over 1+(u+x)^2}+{8x(x-u) \over 
1+(u-x)^2} \bigg] +{3u \over x^2}(3u^4+2u^2-1) \bigg\}  \Bigg\} \nonumber \\ &+& 
{g_A^2 c_4 m_\pi^4 \over (4\pi f_\pi)^4} \Bigg\{{7\over 6u^2}-{317 u^2 \over 54}
+{80u^4 \over 9}+{21+340u^2 \over 27(1+4u^2)} \nonumber \\ && +\bigg[ {16 \over 9u} 
\ln(1+4u^2) -4u^3 -{10 u\over 3}- {64u \over 9(1+4u^2)} \bigg] \arctan 2u   
\nonumber \\ && +\bigg( {47 \over 72}-{7\over 12u^4}-{14 \over 9u^2}+{311u^2 \over 
54} +{16 \over 9(1+4u^2)}\bigg)\ln(1+4u^2)\nonumber \\ && +\bigg({7 \over 96u^6} 
+{49 \over 144u^4}-{1\over 3u^2}-{1\over 3}\bigg)\ln^2(1+4u^2) +\int_0^u\!\!dx 
\bigg\{{L^2 \over 3u}\bigg[{3\over x^2}\nonumber \\ && \times(1+u^2)^3(1-3u^2)+4
(6u^6+7u^4-1)-2x^2(7+9u^4)-4x^4+3x^6 \bigg]\nonumber \\ && +2L \bigg[{7\over 3}
(1-2u^2-3u^4)+{(1+u^2)^2\over x^2}(3u^2-1)\bigg] +{u \over x^2}(1-2u^2-3u^4) \bigg\} 
\Bigg\}  \,,\end{eqnarray}
with the auxiliary function:
\begin{equation} L(x,u) = {1\over 4x}\ln{1+(u+x)^2\over 1+(u-x)^2} \,.\end{equation}
\begin{eqnarray} G_d(\rho)&=&{g_A^2 m_\pi\over 3\pi^2(4f_\pi)^4}\Bigg\{  c_1\bigg[
{16u \over 1+4u^2} -{10\over u}+\bigg({5\over u^3}+{16u \over  1+4u^2}\bigg) 
\ln(1+4u^2)\nonumber \\ && -{5+8u^2\over 8u^5} \ln^2(1+4u^2)\bigg] +c_3\bigg[
2u+{1\over u}-{3\over u^3}-{12u \over 1+4u^2}\nonumber \\ &&
+\bigg({3+5u^2+5u^4 \over 2u^5}-{8u \over 1+4u^2}\bigg) \ln(1+4u^2)-{3+11u^2+
12u^4 \over 16u^7} \nonumber \\ && \times  \ln^2(1+4u^2)\bigg] +c_4 \bigg[
{3\over 2u^3}-{3\over u}+{4u \over 1+4u^2} \nonumber \\ && +{3\over 4u^5}(2u^4-1)
\ln(1+4u^2)+{3+6u^2-8u^4\over 32u^7}\ln^2(1+4u^2)\bigg] \Bigg\} \,,\end{eqnarray}
\begin{eqnarray}G_{so}(\rho) &=& {g_A^2 m_\pi \over \pi^2 (4 f_\pi u)^4}\Bigg\{ 
c_1\bigg[{3+26u^2 +48u^4\over 4u^3} \ln(1+4u^2)-14u^3-10u-{3\over 2u} \nonumber \\
&& -{3+32u^2+80u^4 \over 32 u^5} \ln^2(1+4u^2)\bigg] + c_3 \bigg[{17u^3 \over 3}
-{8 u^5\over 9} -{31u \over 12}\nonumber \\ && -{5 \over 2u}-{5\over 16u^3}+\bigg(
{5\over  32u^5}+{25 \over 16u^3}+{43\over 12u}-{u\over 2}-2u^3\bigg)\ln(1+4u^2)
\nonumber \\ && -{5+60u^2+208u^4+192u^6\over 256 u^7}  \ln^2(1+4u^2)\bigg]+ c_4 
\bigg[{16 u^5\over 9}- {u^3 \over 3} -{7u \over 12}\nonumber \\ && -{1 \over u}
-{5\over 16u^3}+\bigg({5\over  32u^5}+{13 \over 16u^3}+{13\over 12u}+{u\over 2}-
{2u^3\over 3}\bigg)\ln(1+4u^2)\nonumber \\ && -{5+36u^2+80u^4+64u^6\over 256 u^7}  
\ln^2(1+4u^2)\bigg] \Bigg\}\,,\end{eqnarray}
\begin{eqnarray}G_J(\rho) &=& {3g_A^2 c_1 m_\pi \over \pi^2 (4 f_\pi u)^4}\Bigg\{ 
2 u^3 +{33u\over 8}+{1\over 2u} - {8+37u^2+100u^4 \over 32 u^3}
\ln(1+4u^2) \nonumber \\ && -{3\over 2}\arctan 2u +{1+4u^2 \over 32 u^5}\ln^2
(1+4u^2) + 3\int_0^u\!\!dx\bigg\{ {L^2 \over u^2}\bigg[{3\over 4x^2} (1+u^2)^4
\nonumber \\ && +(1+u^2)(1-u^4)+{11 x^6\over 4} +5(1-u^2)x^4 +{x^2\over 2} 
(5u^4-14u^2+5)\bigg]\nonumber \\ && +{L\over 2u} \bigg[3u^4+2u^2-1-{3\over  x^2}
(1+u^2)^3\bigg] +{3\over 4x^2}(1+u^2)^2 \bigg\} \Bigg\} \nonumber \\ &+&
{g_A^2 c_3 m_\pi \over \pi^2 (8 f_\pi u)^4}\Bigg\{ \Big[149-61u^2-102u^4-8u^{-2}  
\ln(1+4u^2)\Big]\arctan 2u \nonumber \\ && +{1216u^5 \over 5}+{875u^3\over 12}
-{303u\over 4}+{4\over u}+{3\over u^3} +{3+16u^2+48u^4 \over 16 u^7} \nonumber
\\ &&  \times  \ln^2(1+4u^2) + \bigg({1687u \over 48}-{45u^3 \over 4}-{309 \over 
16u}- {5\over u^3}- {3\over  2u^5}\bigg) \ln(1+4u^2)\nonumber \\ && 
+3\int_0^u\!\!dx\,\bigg\{{3L^2 \over 2u^2}\bigg[{5\over x^4}(1+u^2)^6 +{6\over x^2}
(1+u^2)^4(1-3u^2) +(1+u^2)^2 \nonumber \\ && \times (23-18u^2+39u^4) + 4x^2(9+23
u^2-5u^4-19u^6) +17x^8 \nonumber \\ && +x^4(19-26u^2+99u^4) +22x^6(1-3u^2)\bigg]
+{L \over u}\bigg[-{15\over  x^4}(1+u^2)^5\nonumber \\ && +{1\over  x^2}(1+u^2)^3
(49u^2-3) -6(17u^6+13u^4+7u^2+11) \bigg] \nonumber \\ && +{15\over  2x^4}
(1+u^2)^4 -{2\over  x^2}(1+u^2)^2(3+11u^2) \bigg\}\Bigg\}\nonumber \\  &+&
{g_A^2 c_4 m_\pi \over \pi^2 (8 f_\pi u)^4}\Bigg\{ \Big[10u^4+95u^2-79+16u^{-2}  
\ln(1+4u^2)\Big]\arctan 2u \nonumber \\ && +{512u^5 \over 15}-{2185u^3\over 12}
+{181u\over 4}+{4\over u}+{3\over u^3} +{3+16u^2-48u^4 \over 16 u^7} \nonumber
\\ &&  \times  \ln^2(1+4u^2) + \bigg( {119 \over 16u}-{3\over  2u^5} -{5\over u^3}-
{173u \over 48}-{9u^3 \over  4}\bigg) \ln(1+4u^2)\nonumber \\ && 
+\int_0^u\!\!dx\,\bigg\{{3L^2 \over 2u^2}\bigg[-{5\over x^4}(1+u^2)^6 +{6\over x^2}
(1+u^2)^4(3u^2-1) -(1+u^2)^2 \nonumber \\ && \times (7+14u^2+23u^4) + 4x^2
(3u^6+5u^4-7u^2-9) -x^8 \nonumber \\ && +x^4(26u^2-3u^4-51) +x^6(2u^2-22)\bigg]
+{L \over u}\bigg[{15\over  x^4}(1+u^2)^5\nonumber \\ && +{1\over  x^2}(1+u^2)^3
(3-49u^2) +18(1+3u^2)(1+u^2)^2 \bigg] \nonumber \\ && -{15\over  2x^4}(1+u^2)^4 
+{2\over  x^2}(1+u^2)^2(3+11u^2) \bigg\}\Bigg\} \,,\end{eqnarray}
with $L(x,u)$ given in eq.(29). A good check of all formulas collected in this section 
is provided by their Taylor-expansion in $k_f$. Despite the superficial opposite  
appearance the leading term in the $k_f$-expansion is $k_f^3$. In several cases 
it is even a higher power of $k_f$. The full Taylor series in $k_f$ has however a small 
radius of convergence $k_f<m_\pi/2$, corresponding to tiny densities $\rho<0.003\,
$fm$^{-3}$. Let us also take the occasion to correct the 
expression for $G_{so}(\rho)$ written in eq.(27) of ref.\cite{isofun}. The correct 
expression is obtained by inserting into eq.(31) the parameters $c_1=0$, $c_3=-g_A^2/2
\Delta$ and $c_4 = g_A^2/4\Delta$. The term omitted in ref.\cite{isofun} vanishes 
(accidentally) for the original density-matrix expansion of Negele and Vautherin 
\cite{negele} but not for the improved density-matrix expansion of Gebremariam, Duguet 
and Bogner \cite{dmeimprov}. The numerical consequences of this correction are 
insignificant since (at half nuclear matter density $\rho_0/2 = 0.08\,$fm$^{-3}$) the 
already small contribution $G_{so}(\rho_0/2) = 6.19$\,MeVfm$^5$ gets just further 
reduced to $G_{so}(\rho_0/2) = 2.63$\,MeVfm$^5$. 
\section{Results and discussion}
In this section we present and discuss our numerical results obtained by summing the 
series of two- and three-body contributions given in sections 3 and 4. The physical 
input parameters are: $g_A=1.3$ (nucleon axial vector coupling constant), 
$f_\pi= 92.4\,$MeV (pion decay constant) and $m_\pi=138\,$MeV (average pion mass).
We use consistently the same parameters pertinent to the chiral three-nucleon 
interaction: $c_E=-0.625$, $c_D=-2.06$, $\Lambda_\chi=700\,$MeV, $c_1=-0.76\,$GeV$^{-1}$, 
$c_3=-4.78\,$GeV$^{-1}$ and $c_4=3.96\,$GeV$^{-1}$, as in our previous work 
\cite{chiral23fun} on the isoscalar part of the nuclear energy density functional.
Let us remind that the low-energy constants $c_E = -0.625$ and $c_D = -2.06$ have been 
obtained in refs.\cite{achim,3bodycalc} by fitting them simultaneously (after fixing 
$\Lambda_\chi=700\,$MeV) to the binding energies of $^3$H and $^4$He using the 
low-momentum NN-interaction $V_{\rm low-k}$ \cite{vlowkreview} at a cutoff scale of 
$\Lambda=414$\,MeV.
\begin{figure}
\begin{center}
\includegraphics[scale=0.41,clip]{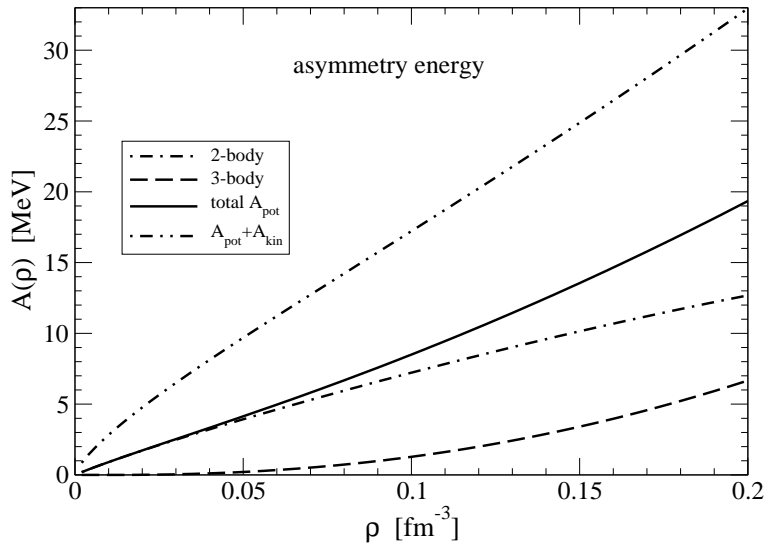}
\end{center}
\vspace{-.9cm}
\caption{Contributions to the asymmetry energy $A(\rho)$ of nuclear matter.}
\end{figure}

Fig.\,3 shows the contributions to the asymmetry energy $A(\rho)$ of infinite
spin-saturated nuclear matter for densities up to $\rho = 0.2\,$fm$^{-3}$. The 
dash-dotted and dashed line give the two-body and three-body contributions to this 
quantity. Their sum, the  total interaction contribution, is shown by the full line 
in Fig.\,3. In the Hartree-Fock approximation the asymmetry energy $A(\rho)$ is 
completed by adding the (relativistically improved) kinetic energy contribution 
$A_{\rm kin}(\rho) =k_f^2/6M-k_f^4/12M^3$, with $M=939\,$MeV the (free) nucleon mass. 
Adding these three pieces together, one obtains for the asymmetry energy at nuclear 
matter saturation density $\rho_0= 0.16\,$fm$^{-3}$ the value $A(\rho_0) = 26.5\,$MeV. 
This is compatible with the empirical values  $A(\rho_0) = (35\pm 2)\,$MeV extracted 
in extensive fits of nuclide masses in refs.\cite{seeger,blaizot}. For comparison, 
a recent microscopic estimate of the asymmetry energy in a relativistic mean-field 
model (constrained by some specific properties of certain nuclei) gave the value 
$A(\rho_0) = (34\pm 2)\,$MeV \cite{dario}. Note that about 1/3 of the empirical value 
$A(\rho_0)$  is provided by the kinetic energy: $A_{\rm kin}(\rho_0)=11.8\,$MeV. 
Another quantity of interest is the slope of the 
asymmetry energy at saturation density. We find for the slope parameter $L = 3\rho_0 
A'(\rho_0) =76\,$MeV which is again compatible with the value $L \simeq 100\,$MeV 
quoted in ref.\cite{blaizot}. It is remarkable that the Hartree-Fock approximation 
works already reasonably well for the asymmetry energy  $A(\rho)$ when giving results 
that are about $20\%$ smaller than empirical determinations. In contrast to this, the 
Hartree-Fock approximation gives a much too shallow binding minimum in the 
equation of state $\bar E(\rho)$ of isospin-symmetric nuclear matter (see Fig.\,7 in 
ref.\cite{chiral23fun}), and second order corrections are very important in order 
to converge eventually to the empirical saturation point \cite{achim,hebeler}. The 
role of three-nucleon forces is also different for both quantities. On the one hand 
side repulsive three-body effects are essential in order to achieve saturation of 
nuclear matter but they do contribute little to the asymmetry energy $A(\rho_0)$, in 
the present calculation merely $4.0\,$MeV. It is also interesting to remind that the 
one-pion exchange alone produces a negative contribution to the asymmetry energy:
\begin{equation}   \tilde A(\rho)^{(1\pi)}= {g_A^2m_\pi^3\over (4\pi f_\pi)^2 }
\bigg\{\bigg({u\over 3}+{1\over 8u}\bigg)\ln(1+4u^2) -{u^3 \over 3}-{u\over 2} 
\bigg\}\,,\end{equation}
which amounts to $-4.4\,$MeV at normal nuclear matter density $\rho_0= 0.16\,$fm$^{-3}$.

\begin{figure}
\begin{center}
\includegraphics[scale=0.41,clip]{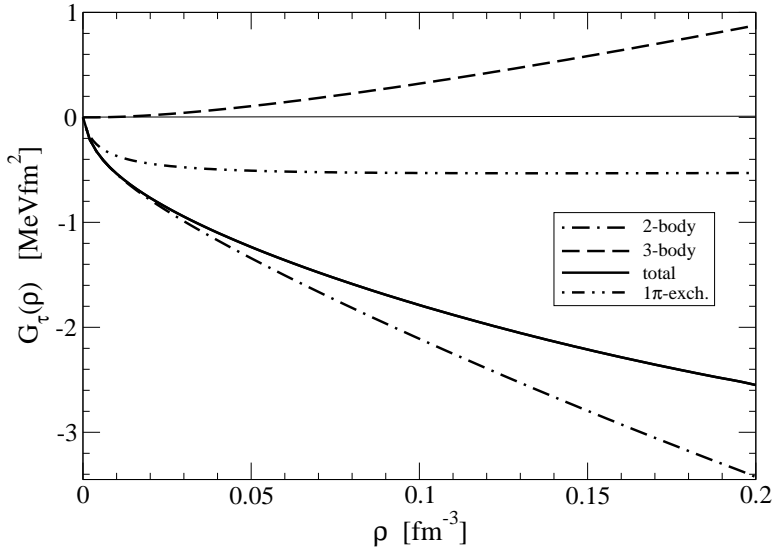}
\end{center}
\vspace{-.9cm}
\caption{Contributions to the strength function $G_\tau(\rho)$ versus the nuclear 
density $\rho$.}
\end{figure}

Fig.\,4 shows the contributions to the strength function $G_\tau(\rho)$. One observes 
that the (negative) two-body contributions get somewhat reduced in size by the 
(positive) three-body corrections. For orientation we have included in Fig.\,4 also 
the one-pion exchange contribution to $G_\tau(\rho)$ as given by the expression:
\begin{equation} G_\tau(\rho)^{(1\pi)}= {g_A^2m_\pi \over 3(4\pi f_\pi)^2}\bigg\{
{u\over 1+4u^2}-{1\over 4u}\ln(1+4u^2) \bigg\}\,, \end{equation}
with $u= k_f/m_\pi$. One recognizes that this relatively small contribution becomes 
almost independent of density for $\rho>0.02\,$fm$^{-3}$. Note that according its 
construction the $G_\tau(\rho)$ term in the nuclear energy density functional eq.(2) 
splits the effective (in-medium) masses of protons and neutrons in linear proportion 
to a (local) isospin-asymmetry $(\rho_p-\rho_n)/\rho$. By comparison
with the isoscalar strength function $F_\tau(\rho)$ (see Fig.\,8 in 
ref.\cite{chiral23fun}) one concludes that the isovector strength function 
$G_\tau(\rho)$ is suppressed by about a factor 5 and of opposite sign. 

Next, we show in Fig.\,5 the strength function $G_\nabla(\rho)$ of the isovector 
surface term $(\vec \nabla \rho_p-\vec \nabla \rho_n)^2$. The three-body contribution 
to this quantity is negligible. Furthermore, one observes from Fig.\,5 
that both components $G_d(\rho)$ and $G_\tau(\rho)/4\rho$ (see eq.(3)) are of equal 
importance for the strength function $G_\nabla(\rho)$. The pronounced decrease at very 
low densities $\rho<0.01\,$fm$^{-3}$ is caused by the $1\pi$-exchange and has also been 
observed in other calculations \cite{microefun}. In phenomenological Skyrme 
parameterizations the strength function $G_\nabla(\rho)^{(\rm Sk)}=-[3t_1(2x_1+1)+t_2
(2x_2+1)]/64$ is a constant, whose value is however not well determined. Taking 
the modern Sly forces \cite{sly} as a guideline one obtains the band $G_\nabla^{(\rm Sk)} 
=-(11\pm 5)\,$MeVfm$^5$ which covers well the results of present (microscopic) 
calculation. For comparison the strength function $F_\nabla(\rho)$ of the isoscalar 
surface term $(\vec \nabla \rho)^2$ is about one order of magnitude larger and 
empirically much better determined: $F_\nabla^{(\rm Sk)} \simeq 75\,$MeVfm$^5$ \cite{sly}.

\begin{figure}
\begin{center}
\includegraphics[scale=0.41,clip]{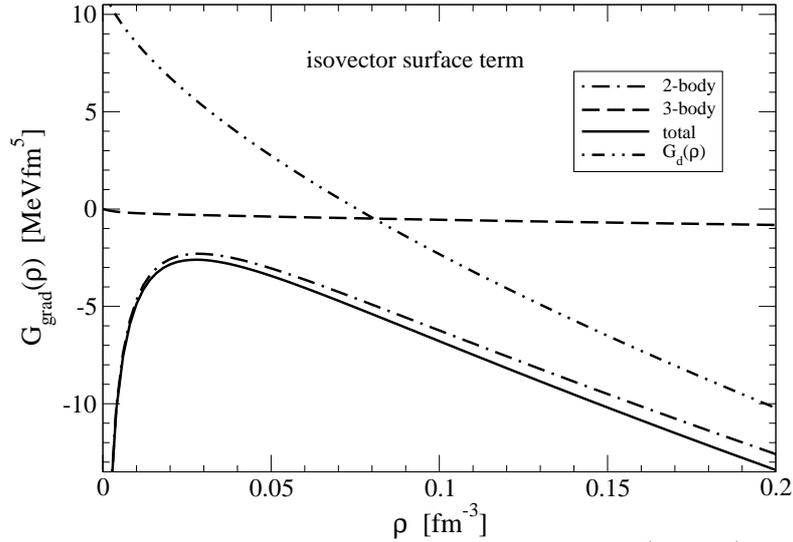}
\end{center}
\vspace{-.9cm}
\caption{Strength function $G_\nabla(\rho)$ of the isovector surface term 
$(\vec \nabla \rho_p-\vec \nabla \rho_n)^2$ versus the nuclear density $\rho$.}
\end{figure}  
\begin{figure}
\begin{center}
\includegraphics[scale=0.41,clip]{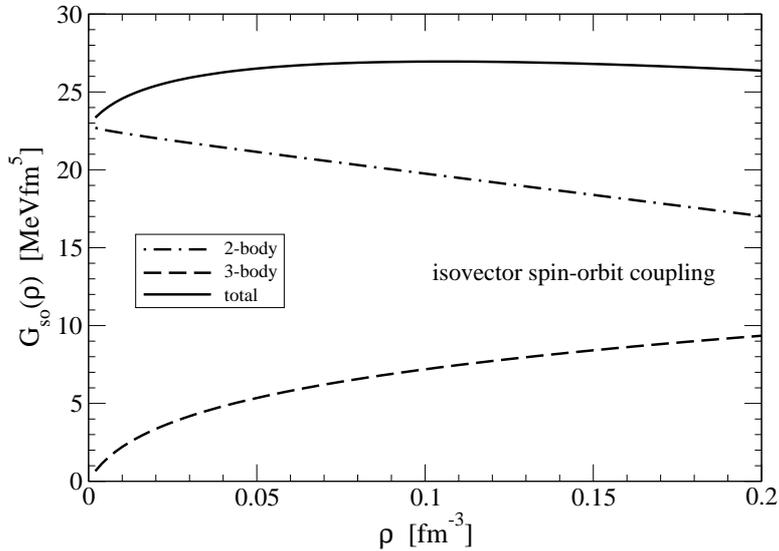}
\end{center}
\vspace{-.9cm}
\caption{Strength function $G_{so}(\rho)$ of the isovector spin-orbit coupling term 
$(\vec \nabla \rho_p-\vec \nabla \rho_n)\cdot (\vec J_p- \vec J_n)$ versus the nuclear 
density $\rho$.}
\end{figure}
\begin{figure}
\begin{center}
\includegraphics[scale=0.41,clip]{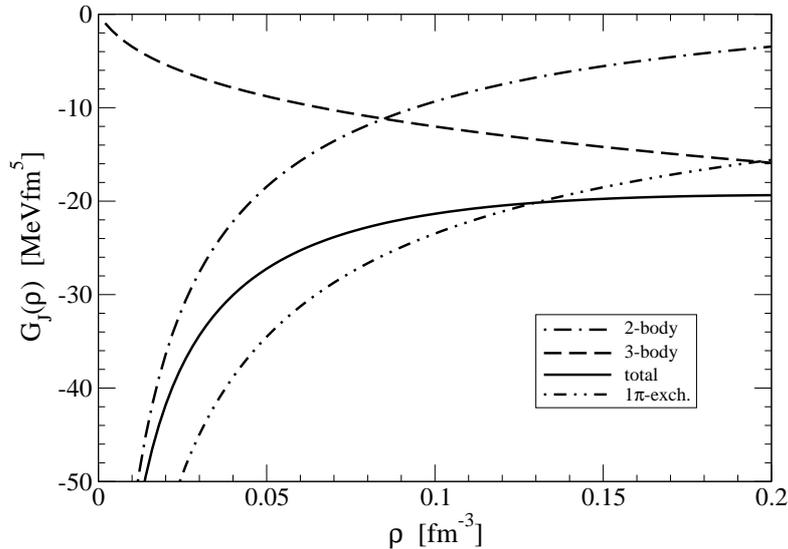}
\end{center}
\vspace{-.9cm}
\caption{Strength function $G_J(\rho)$ multiplying the squared isovector spin-orbit 
density $(\vec J_p-\vec J_n)^2$ versus the nuclear density $\rho$.}
\end{figure}

Next, we show in Fig.\,6 the strength function $G_{so}(\rho)$ of the isovector 
spin-orbit coupling term $(\vec \nabla \rho_p-\vec \nabla \rho_n)\cdot (\vec J_p-
\vec J_n)$. One sees that the weak decrease of the two-body contribution with  
density $\rho$ gets compensated by a small (positive) three-body contribution. The 
resulting total spin-orbit coupling strength $G_{so}(\rho)$ comes out close to the
constant value $G_{so}^{(\rm Sk)}=W_0/4 \simeq 30\,$MeVfm$^5$ of Skyrme parameterizations 
\cite{sly,pearson}.  At this point it should be emphasized that the isovector 
spin-orbit coupling strength in nuclei is presently not well determined. For example, 
no definite choice could be made  in ref.\cite{flocard} between different 
density-dependences $(\sim \rho_p+\gamma  \rho_n,\, \gamma=0,1,2)$ of the neutron 
spin-orbit potential.

Finally, we show in Fig.\,7 the strength function $G_J(\rho)$ accompanying the squared 
isovector spin-orbit density $(\vec J_p-\vec J_n)^2$ in the nuclear energy density 
functional. The two-body and three-body contributions come with equal sign but exhibit 
an opposite density-dependence. For orientation we have included in Fig.\,7 also 
the one-pion exchange contribution to $G_J(\rho)$ as given by the expression:
\begin{equation} G_J(\rho)^{(1\pi)}= {3g_A^2\over (32 m_\pi f_\pi)^2 u^6} \Big[4u^2
-8u^4-\ln(1+4u^2) \Big]\,.\end{equation}
Apparently, the strength function $G_J(\rho)$ is dominated by this unique long-range 
contribution which is also responsible for the strong density-dependence 
\cite{microefun} below $\rho<0.05\,$fm$^{-3}$. At this point it should be kept in mind 
that the  $(\vec J_p-\vec J_n)^2$ term in the nuclear energy density functional eq.(2)
represents non-local Fock contributions from tensor forces etc. An outstanding 
$1\pi$-exchange contribution to $G_J(\rho)$ is therefore not surprising.

In summary we have calculated the strength functions of isovector terms in the nuclear 
energy density functional from chiral two-and three-nucleon interactions. The results
for asymmetry energy $A(\rho)$ suggest that the Hartree-Fock approximation could work 
better for isovector quantities. Clearly, it remains a challenge to confirm this in a 
consistent second-order calculation of the complete nuclear energy density functional. 
\subsection*{Acknowledgement}
I thank J.W. Holt for providing the N$^3$LOW chiral NN-potential in parameterized 
numerical form and for informative discussions.

\end{document}